# Current Control of Magnetic Anisotropy via Stress in a Ferromagnetic Metal Waveguide

Kyongmo An[1], Xin Ma[1], Chi-Feng Pai[3], Jusang Yang[1], Kevin S. Olsson[1], James L. Erskine[1], Daniel C. Ralph[3,4], Robert A. Buhrman[3] and Xiaoqin Li[1,2,*]

[1]Department of Physics, The University of Texas at Austin, Austin, Texas 78712, USA
[2]Center for Complex Quantum Systems, The University of Texas at Austin, Austin, Texas 78712, USA
[3]Cornell University, Ithaca, New york 14853, USA
[4]Kavli Institute at Cornell, Cornell University, Ithaca, New York 14853, USA

*Email address: elaineli@physics.utexas.edu

We demonstrate that in-plane charge current can effectively control the spin precession resonance in an $Al_2O_3$/CoFeB/Ta heterostructure. Brillouin Light Scattering (BLS) was used to detect the ferromagnetic resonance field under microwave excitation of spin waves at fixed frequencies. The current control of spin precession resonance originates from modification of the in-plane uniaxial magnetic anisotropy field $H_k$, which changes symmetrically with respect to the current direction. Numerical simulation suggests that the anisotropic stress introduced by Joule heating plays an important role in controlling $H_k$. These results provide new insights into current manipulation of magnetic properties and have broad implications for spintronic devices.

## I. INTRODUCTION

Magnetic anisotropy plays an important role in the performance of high-density spintronic devices including spin valves[1,2], magnetic tunnel junctions[3-6], and emerging multi-ferroic technologies[7]. Such anisotropy defines the low-energy orientation of the magnetization as well as the stability of the magnetization with respect to external fields, electric currents[8], and temperature-induced fluctuations[9,10]. The control of magnetic anisotropy is typically realized by controlling the growth condition of the magnetic layer[11], switching substrates[12], applying external stress[13], heating[11], or an external electric field[14]. Recently, perpendicular magnetic anisotropy has been achieved in oxide/ferromagnetic metal (FM) heterostructures such as MgO/CoFeB, leading to low critical currents for spin transfer torque switching of tunnel junctions[6]. Therefore, approaches to effectively control magnetic anisotropy as well as elucidating their physical origins become important for further development of multi-functional spintronic devices.

Charge current has recently been utilized to manipulate magnetization including control of magnetic domain wall motions and magnetization switching[3, 15-19]. Efficient control can be achieved using spin-orbit torques (SOTs) originating from either the spin Hall effect in the bulk of a heavy metal[20] or the Rashba effect at a magnetic interface[21]. CoFeB-based alloys have attracted great attention due to their high magneto-resistance[22] and they are commonly used as the electrode material for magnetic tunnel junctions. Although charge-current-induced magnetization manipulation of CoFeB has been extensively studied, current-induced magneto-elastic effects have been rarely discussed, even though CoFeB is known to exhibit a large magneto elastic constant[23].

In this letter, we investigate current-induced magnetic resonance shifts in a CoFeB/Ta waveguide deposited on an $Al_2O_3$ substrate with the Brillouin light scattering (BLS) technique. The magnetic resonance shift exhibits both symmetric and asymmetric dependences when the direction of the direct current (DC) is reversed. A number of mechanisms which can contribute to the asymmetric shift have been investigated previously[21,24], including the Oersted field, the spin Hall effect, and the Rashba effect. In this paper, we focus on the symmetric frequency shift, which can be understood as arising from a current-induced change in the in-plane uniaxial magnetic anisotropy field $H_k$. A modification of $H_k$ up to ~20% is realized using a moderate current of $4 \times 10^6$ A/cm². Numerical simulations suggest that the current-controlled magnetic anisotropy originates at least in part from anisotropic stress in the waveguide, generated by Joule heating from the in-plane current flow. Our study shows that the effective $H$ field induced by anisotropic stress can play an important role in magnetization control in addition to the frequently discussed field-like SOT from the spin Hall effect or interfacial Rashba torque in CoFeB/Ta bilayer structure[25].

## II. SAMPLE STRUCTURE AND CHARACTERIZATION WITH MOKE

The samples investigated are a series of

$Co_{40}Fe_{40}B_{20}(10)/Ta(10)$ films deposited onto an $Al_2O_3$ substrate by sputtering[20], where the numbers in parentheses represent the layer thicknesses in nanometers. Following deposition, the bilayer structure was patterned into a 10-μm-wide and 200-μm-long waveguide. After the deposition of 240-nm-thick $SiO_2$ insulating layer, a 5-μm wide Cu(150)/Au(10) antenna was created on top of the bilayer waveguide, as depicted in Fig. 1(a). From the measured resistance of the bilayer structure, 1930 Ω, the resistivity of bilayer structure of 193 μΩ cm was calculated. These bilayer structures have been previously used to investigate magnetic switching[20] and spin wave amplification via SOTs[26]. While phenomena driven by SOT were observed in this sample, it does not appear to be the most critical mechanism behind the experimental observation of resonance field shifts discussed in this manuscript.

We first characterize the CoFeB samples with magneto optical Kerr effect (MOKE) measurements at room temperature, as presented in Figs. 1(b, c). Due to the strong demagnetization field, the magnetization lies in the *x-y* plane, i.e., the plane of the film. Angle resolved MOKE measurements show that there is in-plane anisotropy. The in-plane easy axis lies along the waveguide $\phi = 0°$ (parallel to the waveguide axis) while the in-plane hard axis is perpendicular to the waveguide at $\phi = 90°$ as shown in Fig. 1(b). The normalized remanent magnetization ($M_r/M_s$) plotted as a function of $\phi$ in Fig. 1(c) confirms that the in-plane magnetic anisotropy is indeed uniaxial. To calculate the uniaxial anisotropy field, $H_k$, we integrated the curve at $\phi = 90°$ in Fig. 1(b), when the magnetic field is applied along the in-plane hard axis:[27]

$$H_k = 2\int_0^1 dm\, H(m), \quad (1)$$

from which we found $H_k = 44 \pm 3$ Oe, where $m = M/M_s$ is the normalized projection of magnetization $M$ along the external field $H$, and $H(m)$ denotes the required external magnetic field to induce the fractional magnetization $m$.

### III. BLS EXPERIMENTS

BLS measurements were then performed to investigate spin waves in the geometry depicted in Fig. 1(a). Because the external magnetic field $H$ is much larger than the saturation magnetic field ~44 Oe obtained from MOKE, the magnetization is kept aligned with the external magnetic field $H$ in our experiments. Damon–Eshbach spin wave modes[28] propagating perpendicular to the magnetization direction were excited by a microwave current through the antenna. A linearly-polarized laser beam was normally incident on the sample surface, and the orthogonal-polarized component of the backscattered light was collected and sent to a Sandercock-type multipass tandem Fabry-Perot interferometer. Fig. 1(d) inset shows a typical BLS raw spectrum from the spin waves propagating along the CoFeB waveguide with a microwave excitation at *f* = 8 GHz. The peak positions of the measured Stokes and anti-Stokes peaks are determined by the microwave source while the linewidth is limited by the frequency resolution of the interferometer. Thus, very limited information can be obtained from the raw BLS spectrum. In the following, we vary the magnitude of the applied magnetic field and the DC to investigate how the DC can modify the magnetic properties of the waveguide.

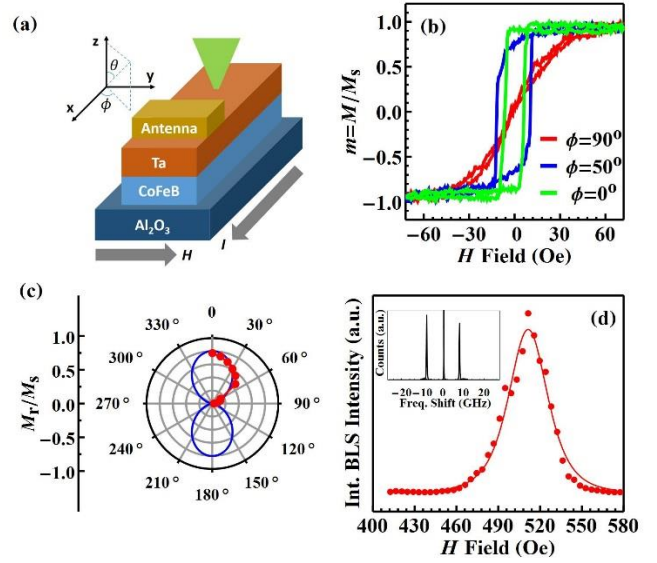

Fig. 1. (a) Schematic illustration of sample geometry used in the BLS experiment. (b) Measured MOKE data with three different magnetic field directions. (c) Polar plot of the normalized remanent magnetization. The solid line shows a cosine function fit of the data. (d) Integrated BLS intensity as a function of external field $H$, where the line is a Lorentzian fit. The inset is the raw BLS spectrum in frequency domain under microwave excitation at a fixed frequency.

To begin, we study how the spin wave intensity, proportional to the integrated BLS intensity, changes with the applied magnetic field at zero DC. The spin wave excited by a fixed microwave frequency exhibits a resonance behavior as shown in Fig. 1(d). The resonance can be well-fitted with a Lorentzian function, from which the peak position $H = H_R$, or the field corresponding to the maximal BLS intensity can be extracted. The resonance field and the frequency of uniform precession can be related by the Smit-Suhl equation[29, 30].

$$f = \frac{\gamma}{2\pi}\sqrt{(H_R - H_k)(H_R + 4\pi M_{eff})}, \quad (2)$$

where $\gamma$ is the gyromagnetic ratio and $4\pi M_{\text{eff}}$ is the effective demagnetization field which also includes the out-of-plane anisotropy field. Strictly speaking, our BLS experiments measure spin waves with small but finite wave vectors instead of the spatially uniform precession. This would lead to a constant offset of $H_R$ by ~ 3% from the peak in BLS-resonance curve, as demonstrated by our previous work on CoFeB/Ta on Si substrates[26]. Because this offset is small, we will approximately equate $H_R$ with the field corresponding to the peak in the BLS spectra as shown in Fig. 1(d).

We then investigate how the resonant magnetic field $H_R$ changes as a DC passes through the waveguide. Our key finding is that $H_R$ decreases with increasing DC as shown in Fig. 2(a) at $f = 8$ GHz. The change in $H_R$ exhibits both symmetric and anti-symmetric behaviors with respect to the DC. The anti-symmetric component can be attributed to a combination of Oersted field, spin Hall effect, and Rashba effect[21,24]. The induced magnetic field from these effects lies along the direction of the external magnetic field, and the direction of the effective field is reversed by reversing the DC direction, leading to anti-symmetric change in $H_R$ with DC.

We focus here on the symmetric reduction of $H_R$ with respect to the DC. Joule heating is known to cause a reduction of $4\pi M_{\text{eff}}$, and hence a symmetric shift in $H_R$. We examine the effect of simple heating by raising the sample temperature uniformly on a heater stage. This control experiment was performed at an excitation microwave frequency of 5 GHz. As shown in Fig. 2(b), $H_R$ is observed to shift upward at a higher temperature, which is opposite to the change in $H_R$ observed in our experiments by passing DCs through the waveguide. Hence, there must exist other mechanisms that overcome the increase of $H_R$ due to the decrease in $4\pi M_{\text{eff}}$ by simple heating and reduce $H_R$ at higher DCs.

To further investigate the origin of the symmetric reduction of $H_R$, $H$ field dependent measurements were performed under different excitation microwave frequencies. The maximal symmetric shift defined by $\Delta H_{\text{symm}}^{\text{m}} \equiv [H_R(I = I_{\max}) + H_R(I = -I_{\max})]/2 - H_R(I = 0)$ is plotted as a function of $H_R(I = 0)$ at each microwave frequency in Fig. 2(c) with a linear fitting line. In other words, $\Delta H_{\text{symm}}^{\text{m}}$ represents the symmetric shift in the resonant field $H_R$ at the highest current ($I_{\max} = 8$ mA) applied in our experiments. To understand the correlation between $\Delta H_{\text{symm}}^{\text{m}}$ and $H_R(I = 0)$, we modify the uniform frequency formula, Eq. (2), to take into account the DC effect phenomenologically as the following

$$f = \frac{\gamma}{2\pi} \sqrt{\begin{aligned}(H_R - H_{k,0} + C_1 I^2)\\ \times (H_R + 4\pi M_{\text{eff},0} + C_2 I^2)\end{aligned}} \quad (3)$$

Here we only keep the lowest-order even contribution from the DC, i.e., the term proportional to $I^2$. $H_{k,0}$ and $4\pi M_{\text{eff},0}$ are the uniaxial anisotropy field and the effective magnetization without DC. The symmetric dependence of $M_{\text{eff}}$ and $H_k$ with respect to DC are explicitly written by introducing $C_1 I^2$ and $C_2 I^2$. With changing DCs, $H_R$ is shifted but $f$ remains the same because of the fixed frequency of the microwave excitation. By taking the derivative with respect to $I^2$, we can obtain the desired relationship[30] between $\Delta H_{\text{symm}}^{\text{m}}$ and $H_R(I = 0)$.

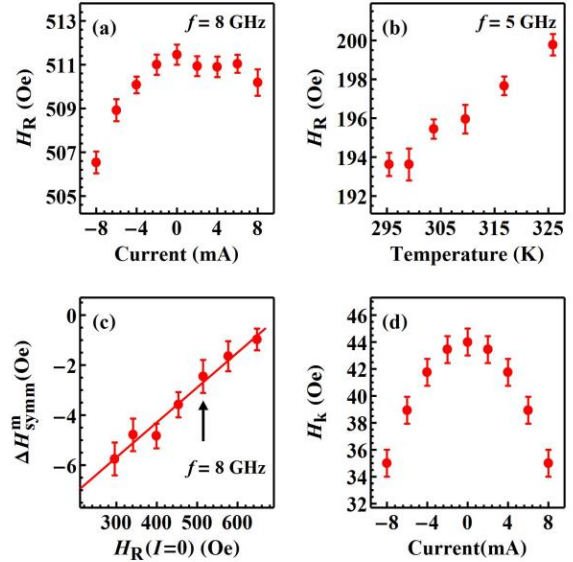

Fig. 2. (a) Measured $H_R$ as a fucntion of DC at $f = 8$ GHz (b) Temperature dependence of $H_R$ at $f = 5$ GHz for uniform heating using a heater stage. (c) The relationship between $\Delta H_{\text{symm}}^{\text{m}}$ and measured $H_R(I = 0)$ at different microwave frequencies, where the solid line is a fit to Eq. (4). The arrow shows the data point at $f = 8$ GHz. Data were taken by varying microwave frequency $f$ in the range of 6-9 GHz with a step size of 0.5 GHz. (d) Current dependence of the uniaxial anisotropy field $H_k$ calculated based on the fitting parameters from Fig. 2(c).

$$\Delta H_{\text{symm}}^{\text{m}} = A_1 H_R(I = 0) + A_2,$$

where,

$$A_1 \equiv -\frac{(C_2 - C_1)I_{\max}^2}{4\pi M_{\text{eff},0}},$$
$$A_2 \equiv -C_1 I_{\max}^2 - A_1(H_{k,0} - C_1 I_{\max}^2). \quad (4)$$

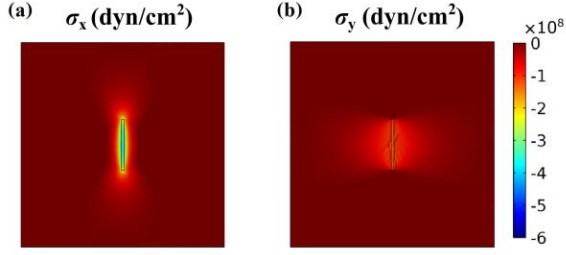

Fig. 3. Calculated stress disctribution along (a) $x$-direction and (b) $y$-direction. The center strip is the CoFeB/Ta waveguide.

Thus, $A_1$ and $A_2$ correspond to the slope and $y$-intercept of the fitting line and are determined to be $0.014 \pm 0.001$ and $-9.9 \pm 0.5$ Oe, respectively. Using these values, we determine $C_1 I_{max}^2 = 9.4 \pm 0.5$ Oe and $C_2 I_{max}^2 = (-0.014 \pm 0.001) \, 4\pi M_{eff}|_{I=0} + 9.4$ Oe. We interpret the $C_2$ term as the reduction of $4\pi M_{eff}$ caused by Joule heating. Based on the Bloch's law[31], ~1.4% reduction of $4\pi M_{eff}$ corresponds to a temperature rise of 22 K. The $C_1$ term can be interpreted as the change in $H_k$, which decreases by 20% at $I = I_{max}$. Based on the $C_1$ and $C_2$ values, we plot $H_k$ as a function of DC using $H_k = H_k|_{I=0} - C_1 I^2$, as shown in Fig. 2(d).

## IV. SIMULATION RESULTS

Next, we explore the possibility that the anisotropic stress, induced by Joule heating from current flow through the bilayer waveguide, plays an important role in the modification of $H_k$. We used the thermal stress module of COMSOL software[30]. We took the power dissipation through the waveguide as a heat source and calculated spatial profiles of stresses. Fig. 3 shows that the calculated stress values for the waveguide along $x$ ($\sigma_x$) and $y$ ($\sigma_y$) directions at $I = 8$ mA. The stress values are negative, indicating that the larger thermal expansion of CoFeB/Ta compared to the $Al_2O_3$ substrate leads to compressive stresses on CoFeB. The anisotropic stresses arise mainly due to the stripe-like shape of the waveguide, as the stress difference between two axes becomes zero if the waveguide has a square rather than rectangular geometry. Based on the volume averaged stress values, we calculated the magneto-elastic energy $E_\sigma$ given by[27]

$$E_\sigma = \frac{3}{2} \lambda \left( \sigma_x \sin^2\phi + \sigma_y \cos^2\phi \right), \quad (5)$$

where $\lambda$ is the magneto-elastic constant of CoFeB, $20 \times 10^{-6}$ [23]. $\phi$ is the angle between $x$ axis and the magnetization as shown in Fig. 1(a). The effective magnetic field associated with $E_\sigma$ can change the uniform frequency formula. By adding the stress induced energy $E_\sigma$ to the total magnetic free energy $E$ and using the Smit-Suhl formula[29, 30], we

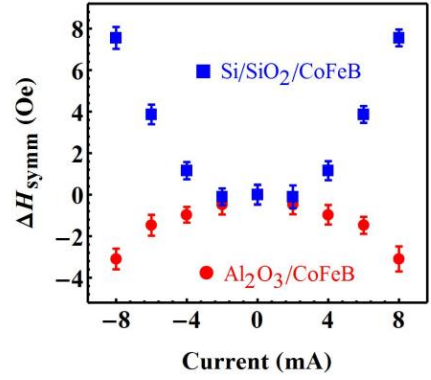

Fig. 4. Measured $\Delta H_{symm}$ as a function of current at 8 GHz microwave frequency for a CoFeB waveguide on $Al_2O_3$ (red) and $Si/SiO_2$ (blue) substrates.

obtain the modified uniform frequency formula given by

$$f = \frac{\gamma}{2\pi} \sqrt{\left( H_R - \left( H_k - \frac{3\lambda}{M_s}(\sigma_x - \sigma_y) \right) \right) \times (H_R + 4\pi M_{eff})} \quad (6)$$

With the calculated stress difference $\sigma_x - \sigma_y = 1.6 \times 10^8$ dyn/cm$^2$ and $M_s = 1273 \pm 80$ emu/cm$^3$ [23], we obtain a stress induced field of $7.5 \pm 0.5$ Oe, which is reasonably close to the measured $H_k$ decrease of 9.5 Oe at $I = \pm 8$ mA.

To further confirm that anisotropic stress plays a key role in the observed magnetic resonance shift with DCs, we compare the observed symmetric change in the resonance field defined by

$$\Delta H_{symm} \equiv \frac{H_R(I) + H_R(-I)}{2} - H_R(0). \quad (7)$$

The data for $Al_2O_3/CoFeB(10)/Ta(10)$ and $Si/SiO_2(500)/CoFeB(10)/Ta(10)$ are shown in Fig. 4. The CoFeB waveguide on the Si substrate was 8 $\mu$m-wide and 270 $\mu$m-long. $\Delta H_{symm}$ for CoFeB on the Si substrate increases with DCs, which is consistent with a simple Joule heating effect while that of CoFeB on $Al_2O_3$ substrate decreases with DCs.

A similar COMSOL calculation was performed for the $Si/SiO_2/CoFeB(10)/Ta(10)$ structure. The calculated stress difference $\sigma_x - \sigma_y$ was only $2.0 \times 10^7$ dyn/cm$^2$. Since $E_\sigma$ depends on the difference in stresses, this leads to a much smaller $\Delta H_{symm}$ compared to the one on the $Al_2O_3$ substrate. This small difference between $\sigma_x$ and $\sigma_y$ originates from the fact that $SiO_2$ has a small thermal expansion coefficient ($0.6 \times 10^{-6}$) compared to that of $Al_2O_3$ ($7.5 \times 10^{-6}$). Thus, the stress from the anisotropic thermal expansion of CoFeB on the Si substrate is limited and the isotropic thermal stress dominates[30].

## V. CONCLUSION

In conclusion, we have investigated the uniaxial magnetic anisotropy field of a CoFeB/Ta waveguide on an $Al_2O_3$ substrate and its dependence on in-plane charge current with the BLS technique. The in-plane uniaxial magnetic anisotropy field is modified by 20% at a modest charge current of $4 \times 10^6$ A/cm$^2$. The modification of $H_k$ is symmetric with respect to the current direction, which cannot be explained by either spin Hall or the Rashba effects. Our simulations suggest that anisotropic stress induced by Joule heating from DCs passing the waveguide can cause a change in $H_k$, which agrees reasonably well with the experimental observation. This Joule heating induced anisotropic stress control of magnetic anisotropy may offer additional design flexibility in the development of new spintronic devices, such as spin valves and magnetic tunneling junctions.

## ACKNOWLEDGEMENTS

The work at UT-Austin (K. An., X. Ma, K. S. Olsson, X. Li) is supported by SHINES, an Energy Frontier Research Center funded by the U.S. Department of Energy (DoE), Office of Science, Basic Energy Science (BES) under award # DE-SC0012670. The work at Cornell was supported by the NSF/MRSEC program (DMR-1120296) through the Cornell Center for Materials Research.

# Supplemental Material: Current Control of Magnetic Anisotropy via Stress in a Ferromagnetic Metal Waveguide

Kyongmo An[1], Xin Ma[1], Chi-Feng Pai[3], Jusang Yang[1], Kevin S. Olsson[1], James L. Erskine[1], Daniel C. Ralph[3,4], Robert A. Buhrman[3] and Xiaoqin Li[1,2,*]

[1]*Department of Physics, University of Texas, Austin, Texas 78712, USA*
[2]*Center for Complex Quantum Systems, The University of Texas at Austin, Austin, Texas 78712, USA*
[3]*Cornell University, Ithaca, New york 14853, USA*
[4]*Kavli Institute at Cornell, Cornell University, Ithaca, New York 14853, USA*

*Email address: elaineli@physics.utexas.edu*

### S1. Derivation of uniform frequency formula

The uniform precession frequency formula can be derived from the Smit-Suhl formula given by

$$f = \frac{\gamma}{2\pi} \frac{1}{M_s \sin\theta} \left[ \frac{\partial^2 E}{\partial \theta^2} \frac{\partial^2 E}{\partial \phi^2} - \left( \frac{\partial^2 E}{\partial \theta \partial \phi} \right)^2 \right], \quad (S1)$$

where $\gamma$ is the gyromagnetic ratio and $M_s$ is the saturation magnetization. $\theta$ and $\phi$ are the angles that represent the magnetization direction defined in Fig. 1(a) of the main text. $E$ is the energy associated with magnetization of our system given by

$$E = -\boldsymbol{H} \cdot \boldsymbol{M} + \frac{1}{2} M_s (4\pi M_{\text{eff}} \cos^2\theta - H_k \sin^2\theta \cos^2\phi), \quad (S2)$$

where $M_{\text{eff}}$ is the effective magnetization and $H_k$ is the in-plane uniaxial anisotropy field. For an in-plane external magnetic field perpendicular to the waveguide, we found numerically the equilibrium direction of magnetization by minimizing the total energy. With the energy term and the calculated equilibrium orientation of magnetization, we can calculate the uniform precession frequency using the Smit–Suhl's formula. Then we obtain

$$f = \frac{\gamma}{2\pi} \sqrt{(H_R - H_k)(H_R + 4\pi M_{\text{eff}})}. \quad (S3)$$

### S2. Derivation of the relationship between $\Delta H_{\text{symm}}^{\text{m}}$ and $H_R$

We take the derivative of Eq. (3) in the main text with respect to $I^2$.

$$\frac{4\pi^2 d(f^2)}{\gamma d(I^2)} = 0 = \left( \frac{dH_R}{dI^2} + C_2 \right)(H_R - H_{k,0} + C_1 I^2) + (H_R + 4\pi M_{\text{eff},0} + C_2 I^2)\left( \frac{dH_R}{dI^2} + C_1 \right). \quad (S4)$$

Further simplifying, we obtain a formula for the change in $H_R$ with respect to $I^2$ given by

$$\begin{aligned}
\frac{dH_R}{dI^2} &= -C_1 - \frac{1}{4\pi M_{\text{eff},0}}(C_2 - C_1)(H_R - H_{k,0} + C_1 I^2) + O\left[ \left( \frac{1}{4\pi M_{\text{eff},0}} \right)^2 \right] \\
&\approx -\frac{1}{4\pi M_{\text{eff},0}}(C_2 - C_1)H_R - C_1 - \frac{1}{4\pi M_{\text{eff},0}}(C_2 - C_1)(-H_{k,0} + C_1 I^2).
\end{aligned} \quad (S5)$$

In the calculation above, we only keep the first order term of $1/(4\pi M_{\text{eff},0})$. $dH_R/dI^2$ can be approximated to $\Delta H_{\text{symm}}^m/I_{\text{max}}^2$ because the change in $H_R$ is observed to be only 1% of $H_R$ when $I$ increases to $I_{\text{max}} = 8$ mA. Then, we obtain

$$\Delta H_{\text{symm}}^m = A_1 H_R + A_2,$$

where,

$$A_1 \equiv -\frac{(C_2 - C_1)I_{\text{max}}^2}{4\pi M_{\text{eff},0}}$$

$$A_2 \equiv -C_1 I_{\text{max}}^2 - \frac{(C_2 - C_1)(-H_{k,0} + C_1 I^2)}{4\pi M_{\text{eff},0}} I_{\text{max}}^2.$$

(S6)

### S3. Comsol calculation

The thermal stress module of COMSOL software was used to calculate the spatial profiles of stresses and strains for both $Al_2O_3/CoFeB(10)/Ta(10)$ and $Si/SiO_2(500)/CoFeB(10)/Ta(10)$ due to the Joule heating at $I = 8$ mA. The size of the waveguide used in the simulation was the same as the actual sample size. The $Al_2O_3$ substrate was assumed to be 8 mm × 8 mm laterally with a thickness of 0.8 mm. The temperature at the bottom of the substrate was fixed at a temperature of 293 K. An adiabatic boundary condition was assumed for other surfaces. We chose a mechanical boundary condition that allows free expansions along all directions. The mesh size was about 3 μm laterally. Five meshes were distributed evenly for each layer along the thickness direction. The calculated volume averaged stresses and strains over the waveguide are shown in Table I. The material properties used in the simulation are summarized in Table II.

Table I. Calculated volume averaged values over the CoFeB layer for the temperature rise, stress, and strain tensors at $I = 8$ mA

| Sample | $\Delta T$(K) | $\sigma_{xx}$ (MPa) | $\sigma_{yy}$ (MPa) | $\varepsilon_{xx}$ | $\varepsilon_{yy}$ |
|---|---|---|---|---|---|
| $Al_2O_3$/CFB/Ta | 21 | - 42 | - 26 | $4.2 \times 10^{-5}$ | $1.7 \times 10^{-4}$ |
| $Si/SiO_2$/CFB/Ta | 31 | - 83 | - 81 | $3.3 \times 10^{-6}$ | $2 \times 10^{-5}$ |

Table II. Parameters used in the comsol simulations

| Material | Thermal conductivity (W/(m·K)) | Thermal expansion coefficient (10$^{-6}$) | Young's modulus (GPa) | Poisson ratio | Heat capacity (J/(kg·K)) | Density (kg/m³) |
|---|---|---|---|---|---|---|
| Ta | 57 | 6.3 | 186 | 0.34 | 140 | 16690 |
| CoFeB | 90 | 12 | 162 | 0.3 | 500 | 8900 |
| $Al_2O_3$ | 30 | 7.5 | 345 | 0.27 | 760 | 3970 |
| Si | 130 | 2.6 | 150 | 0.22 | 700 | 2330 |
| $SiO_2$ | 1.4 | 0.6 | 70 | 0.17 | 700 | 2200 |

### S4. Stress-strain relation

The stress tensor $\sigma_{ij}$ is related with the strain tensor $\varepsilon_{ij}$ by the following equation.

$$\sigma_{ij} = \frac{E}{(1+\nu)(1-2\nu)}\left[(1-2\nu)\varepsilon_{ij} + \sum_k \nu\delta_{ij}\varepsilon_{kk}\right] - \delta_{ij}\frac{E\alpha\Delta T}{(1-2\nu)}, \tag{S7}$$

where $E$ is the Young's modulus, $\nu$ is the Poisson ratio, $\alpha$ is the thermal expansion coefficient, $\delta_{ij}$ is the Kronecker delta, and $\Delta T$ is the temperature change. The equation consists of a strain dependent part (square bracket) and strain independent part. Thus only the first term (strain dependent term) can give rise to the anisotropic stress. For CoFeB on the $Al_2O_3$ substrate, the strain dependent term is comparable with the other term. However, as shown in Table I, CoFeB on the Si substrate has strain values about one order of magnitude smaller compared to CoFeB on $Al_2O_3$ due to the small thermal expansion coefficient of $SiO_2$. Thus, it has a negligible contribution from the strain dependent term and the strain independent term contributes dominantly, leading to the nearly isotropic stress.